**Impact of size effects on photopolymerization and its optical monitoring *in-situ***

Andrea Camposeo,[a,*] Aristein Arkadii,[b] Luigi Romano,[a] Francesca D'Elia,[c] Filippo Fabbri,[a]

Eyal Zussman,[b] Dario Pisignano[a,d]

[a]*NEST, Istituto Nanoscienze-CNR and Scuola Normale Superiore, Piazza S. Silvestro 12, I-56127*

*Pisa, Italy*

[b]*Department of Mechanical Engineering, Technion-Israel Institute of Technology, Haifa, 32000,*

*Israel*

[c]*NEST, Scuola Normale Superiore, Piazza San Silvestro 12, I-56127 Pisa, Italy*

[d]*Dipartimento di Fisica, Università di Pisa, Largo B. Pontecorvo 3, I-56127 Pisa, Italy*

*\*Corresponding author: e-mail: andrea.camposeo@nano.cnr.it*

**Abstract**

Photopolymerization processes are exploited in light exposure-based 3D printing technologies, where either a focused laser beam or a patterned light sheet allows layers of a UV curable, liquid pre-polymer to be solidified. Here we focus on the crucial, though often neglected, role of the layer thickness on photopolymerization. The temporal evolution of polymerization reactions occurring in droplets of acrylate-based oligomers and in photoresist films with varied thickness is investigated by means of an optical system, which is specifically designed for *in-situ* and real-time monitoring. The time needed for complete curing is found to increase as the polymerization volume is decreased below a characteristic threshold that depends on the specific reaction pathway. This behavior is rationalized by modelling the process through a size-dependent polymerization rate. Our study highlights that the formation of photopolymerized networks might be affected by the involved volumes regardless of the specific curing mechanisms, which could play a crucial role in optimizing photocuring-based additive manufacturing.

## 1. Introduction

Photopolymerization processes, which lead to the formation of solid polymer networks starting from monomers and oligomers, are arousing a continuously increasing interest in many fields of science and technology. This is motivated by the importance of predicting the resulting physical properties of polymerized systems with improved accuracy, and of enhancing various applications in surface coatings, adhesives, dentistry, lab on chips, soft actuators and photonics [1-3]. In photopolymerization, light is used for curing pre-polymer solutions which contain suitable photoinitiators. These photosensitive molecules produce reactive species upon photon absorption and initiate reactions proceeding through free-radical or cationic species [4,5]. Such processes show features that make them highly suitable for micro- and nanofabrication for various aspects. First, the intensity profile of a light beam can be structured by using a physical shadow mask, a digital micro-mirror array or a liquid-crystal phase mask [6-8], and such structured light beams can be used to pattern thin layers of photosensitive pre-polymer mixtures. Moreover, light can be focused into diffraction-limited volumes through optical systems, controlling the focal spot size by the numerical aperture (NA) of the focusing optics, thus allowing photopolymerization to be localized at sub-micrometric scale, whereas complex patterns can be realized by scanning the light beam [9]. The absorption of light can occur also by nonlinear phenomena, such as two-photon absorption [10,11], and single or multicolor activation and deactivation processes [12-14], enabling additional options for spatial localization. Technologies impacted by these aspects span from photolithography to laser writing and 3-dimensional (3D) vat photopolymerization [15-18]. Moreover, photopolymerization can be combined with self-assembling, phase-separation and ink-jet processes, which are used, for instance, to fabricate arrays of lenses and micro-structures [19-21].

As the polymerization volume is shrunk down to micrometric and sub-micrometric size, the rate of the process might change. Pioneering studies [22] on planar layers of polymers showed a slowing down of polymerization in thinner films. These findings were attributed to the availability of smaller amounts of monomers and to more effective oxygen diffusion and consequent polymerization quenching [23]. Furthermore, the variation of the polymerization kinetics for pre-polymers in confined volumes has been studied experimentally and modelled through the features of polychromatic kinetics of reactions [24].

Developing an in-depth knowledge of such effects is highly important for photopolymerization-based additive manufacturing, where objects are mostly realized through a layer-by-layer method. Indeed, any volume-dependent variation of the photo-polymerization kinetics would require other process parameters to be correspondingly adjusted. This might be especially critical for those methods where the layer thickness is varied continuously during printing in order to account for different feature sizes. Finally, this is relevant for 3D printing of nanocomposites [25], where photocuring is not spatially-uniform but instead proceeds through possibly reduced, polymer-excess regions. To address these issues, both *in-situ* [26] tools for monitoring photopolymerization and effective models of the polymerized network formation in confined and complex architectures are needed. For instance, *in-situ* and real-time inspection of photopolymerization is possible by measuring the intensity of the luminescence from embedded fluorescent monomers [22], by interferometric measurement of the thickness of the cured layer [27], by infrared spectroscopy [28], by using additional probe lasers [29], by photorheology [30], and by fiber optics sensors [31]. However, using these techniques for monitoring 3D architectures is not straightforward, and attempts for truly closed-loop control of vat photopolymerization processes are being designed [32].

Here, we investigate the temporal evolution of photopolymerization by varying the volume of the pre-polymer. Two widely adopted sample geometries are studied, namely droplets with variable height and thin films with variable thickness. Photopolymerization is monitored *in-situ* and in real-time by measuring the intensity of the same ultraviolet (UV) beam used for curing, which is backscattered by the polymerized region providing micrometric spatial resolution and temporal resolution of a few tens of milliseconds. For both free-radical and cationic/photothermal polymerization, curing times are found to increase by decreasing the volume of the pre-polymer. While the specific mechanisms underneath relate to the particular reaction pathway (which might include oxygen-due inhibition, thickness-dependent heating, and others), the reduction of the polymerization rate upon decreasing thickness appears to be general. These findings are rationalized with a model which introduces a thickness-dependent polymerization rate, well describing the trends measured for the intensity of the backscattered light. Monitoring of polymer photocuring *in-situ* and in real-time, and accounting for size-dependent polymerization, might be relevant for next-generation, intelligent additive manufacturing technologies.

## 2. Material and methods

### 2.1 Materials

Various UV-curable feedstock compounds are used as summarized in Table 1. These materials are selected for investigating different polymerization reactions and timescales, the latter depending on the coefficient of light absorption at 405 nm (the wavelength of the light used for curing). The bisphenol-A-ethoxylate dimethacrylate (BisEMA) oligomer ($M_n$=1700, Sigma Aldrich) is mixed with 2% weight:weight (w:w) of 2,2-dimethoxy-2-phenylacetophenone photoinitiator (Sigma Aldrich, Table 1). The mixture is vortexed until the complete dissolution of the photo-initiator [24] and, afterwards, it is degassed in a vacuum chamber. The absorption spectrum of this pre-polymer mixture is shown in the Supplementary Figure 1. The photoinitiator has a strong absorption band at wavelengths <300 nm and a weaker absorption band in the 300-350 nm range. The latter further extends in the visible range, which allows radiation with a wavelength of 405 nm to be used for photopolymerization. The mixture is deposited on quartz substrates by spin-coating (8000 rpm for 60 s). So-formed films are uniform, but they undergo a dewetting process within a few minutes. This leads to the formation of an ensemble of microdroplets with size ranging from 1 mm down to 1 μm (Supplementary Figure 2).

E-shell® 600 (Envisiontec) is a photocurable resin containing a phosphine oxide photoinitiator [33]. It is used as received by the manufacturer (Table 1). Droplets of E-shell® 600 (Envisiontec) are deposited by material jetting (the set-up is illustrated in the Supplementary Figure 3) on a 2×2 cm² quartz substrate, by using a F5200N.2 robotic system (Fisnar). Droplet jetting of E-shell® 600 is carried out using a gauge 32 needle and by setting a pressure of 8 psi and a deposition time of 0.1 seconds. After printing, the droplets are photopolymerized by using the Micro Plus HD (ENVISIONTEC®) 3D printer (exposure time 1-9 s, intensity of UV light of 3 mW cm$^{-2}$).

SU-8 (Microchem) is an epoxy resin containing triarylsulfonium salts as photoacid generator [34, 35]. It is used as received by the manufacturer without further addition or purification (Table 1). Thin films of SU-8 are deposited on cleaned quartz substrates (1×1 cm$^2$) by spin-coating. The thickness of the films is varied in the range 0.5-5 μm by changing the spin-coating speed (500-8000 rpm) and the used pre-polymer (SU-8 2000.5 for thickness < 1 μm and SU-8 2002 for thickness > 1 μm, respectively, both from Microchem). The deposition of the SU-8 films is followed by a soft-bake step in an oven (~95 °C for 1 hour).

| UV-curable resin | Composition | Type of photopolymerization | Sample morphology | Reference |
|---|---|---|---|---|
| BisEMA mixture | BisEMA oligomer and 2% w:w of 2,2-dimethoxy-2-phenylacetophenone | Free-radical | Droplets | [24] |
| E-shell® 600 | As-received by the manufacturer (Envisiontec) | Free-radical | Droplets | [33] |
| SU-8 | As-received by the manufacturer (MicroChem) | Cationic | Thin films | [34, 35] |

Table 1. Properties of pre-polymers used in the study.

The thickness of the samples is measured by a stylus profilometer (Dektak). The morphology of the printed structures is characterized by scanning electron microscopy (SEM), using a Merlin system (Zeiss).

## 2.2 Photopolymerization and *in-situ* monitoring

The system used for photopolymerization and *in-situ* monitoring of the process is based on a confocal microscope set-up (mod. FV1000, Olympus). The main components of this set-up are schematically illustrated in Figure 1a. It includes a diode laser source with 405 nm emission wavelength and maximum power of 1 mW, as measured at the sample position by using a power meter (mod. 843-R, Newport). The laser is coupled to an inverted microscope (IX71, Olympus) and focused onto the sample by a 10× objective (Numerical Aperture, NA=0.4). The spot size ($2w_0$, where $w_0$ is the beam waist) is about 1.2 μm. This system can be used also for 3D printing experiments through vat photopolymerization, performed by using an approach similar to the one reported in Ref. [36]. While in standard laser 3D printing, low-NA focusing systems with high focal depth (50-100 μm) are used, in the system sketched in Figure 1a the focusing conditions allow the photopolymerization to occur only in a volume close to the focal spot, providing polymerized layers with thickness of a few micrometers. The focal spot can be moved in 3D by shifting the objective position with respect to the sample along the optical axis, and by 2D-scanning of the laser spot in the sample plane (i.e. the plane perpendicular to the optical axis) by a galvo-mirror (Figure 1a). 3D objects can be therefore manufactured by these two independent actuation systems. The laser on and off switching can be controlled with microseconds time resolution through an electro-optic shutter. The 3D structures here realized are printed with a layer thickness of 2 μm, much lower than values (10-100 μm) used in standard UV stereolithography systems [16]. After printing, the uncured pre-polymer is rinsed by isopropanol. The maximum area that can be printed is $1.2 \times 1.2$ mm$^2$.

The monitoring of the photopolymerization process is performed *in-situ* by collecting the photons of the 405 nm laser (the same used for photopolymerization), that are backscattered by the sample. More specifically, the change of the refractive index of the pre-polymer during the photopolymerization process can lead to a variation of the intensity of both the back-reflected light and the light that is backscattered within the collection angle of the focusing objective [37,38]. Therefore, the intensity ($I_{BS}$) of the UV laser beam, focused in the pre-polymer and back-scattered by the polymerized area, is indicative of the local evolution of polymerization. This approach is similar to refractometry experiments [31], with the main advantage of using the same light source for both curing and process sensing.

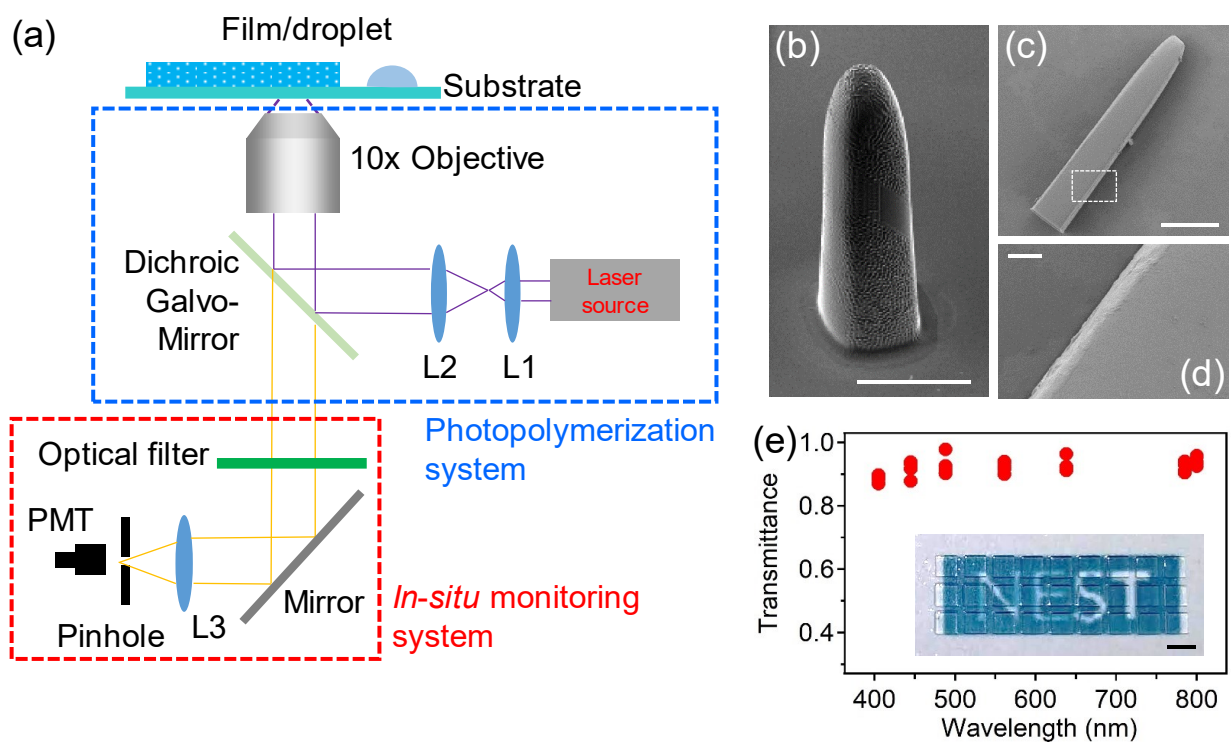

**Figure 1**. (a) Schematic illustration of the experimental set-up used for printing 3D structures by UV laser polymerization, and for concomitantly measuring the intensity of the light backscattered by the polymerized region, here exploited for monitoring *in-situ* the curing processes. L1, L2 and L3: lenses; PMT: photomultiplier. (b)-(d) SEM micrographs of printed vertical structures made of BisEMA. Panel (d) shows a magnification of the vertical edge of the structure highlighted by a dashed square in (c). Scale bars: (b)-(c) 100 μm, (d) 10 μm. (e) Optical transmittance, at various wavelengths, for a flat window (1×1 mm², thickness = 425 μm) printed with the system shown in (a). The inset shows an array of 9×3 of such printed windows, evidencing their good optical transparency in the visible (the logo of the NEST laboratory can be clearly seen through them).

The backscattered light is collected by the same objective used for focusing the light onto the sample (Olympus 10× objective NA=0.4), and directed to a photomultiplier (Figure 1a), utilizing the optics embedded in the confocal microscopy system. The pinhole of the confocal system can be also exploited in order to decrease the depth and diameter of the sample that is probed, and by which backscattered light is collected. This system allows the photopolymerization kinetics occurring in a volume nearby the focal spot of the 405 nm laser to be monitored. The intensity of the 405 nm laser backscattered by the sample is measured with a temporal resolution of a few tens of ms.

## 2.3 Optical properties

Micro-Raman measurements are carried out with a Renishaw InVia spectrometer equipped with a confocal optical microscope, by using a 532 nm laser as excitation source and a 50× long working distance objective with NA= 0.5 for excitation and collection of the scattered signals. Every spectrum is acquired with an excitation power of 5 mW and an acquisition time of 5 s. For E-shell® 600, the conversion factor is calculated as $C_R = 1 - \left(\frac{I_{1638}}{I_{1608}}\right) / \left(\frac{I_{1638}}{I_{1608}}\right)_{ref}$, where $I_{1638}$ and $I_{1608}$ are the intensities of the C=C aromatic ring stretching and the aliphatic C=C stretching modes, respectively (measured by Raman spectroscopy), and the ratio $(I_{1638} / I_{1608})_{ref}$ is calculated for a reference sample [39,40].

The UV and visible spectra of the samples are measured by a Cary 900 (Varian) spectrophotometer. The samples to be investigated are deposited on a quartz substrate and positioned with their surface perpendicular to the incident light beam.

The coefficient of optical transmission of the printed structures at various visible wavelengths is measured by using a system composed by five laser sources coupled to a single mode optical fiber (mod. L6Cc, Oxiuss), a diode laser at 785 nm (Thorlabs) and a Ti-Sapphire (Mira, Coherent, optically pumped by mod. Verdi V10, Coherent). The beam size of the laser sources is of the order of 0.5 mm. The use of monochromatic and collimated light beams allows the intensity attenuation of the light beam passing through the printed samples to be measured. To this aim, all the measurements are performed by placing the samples at normal incidence with respect to the incident light beam and measuring the power of the laser beam before ($P_{in}$) and after passing through the sample ($P_{out}$) with the power meter. Then, the transmittance, $T$, is calculated as the ratio $T=P_{out}/P_{in}$, while the fraction of incident light intensity that is absorbed by the sample, $A$, is obtained as $A=1-T-R$, where the reflected light is estimated by the reflectance coefficient at normal incidence [41]: $R = \left(\frac{n-1}{n+1}\right)^2$, where $n$ is the refractive index of the sample. These measurements are performed with the BisEMA samples, for which, $n$=1.49 [42].

## 3. Experimental results

Figure 1b shows a 3D structure printed by BisEMA, and imaged by SEM. No footprint of the printed layers can be appreciated in the realized object. Such layered footprint, possibly leading to a step-like surface structure [43], could constitute a major flaw, especially when 3D printing is used to realize optical components. Various approaches have been developed to overcome such drawback, such as post-processing by mechanical polishing [44], depositing and curing additional layers [44-46], using micro-continuous liquid interface production [47], or optimized polymer formulations [48]. Here we use a highly focused laser, and very thin printing layers (2 µm), leading to surfaces of the printed structures with submicrometric roughness. This roughness is estimated through the height of the features protruding from the surfaces of the printed structures (<0.5 µm, Figure 1d). Flat windows (area $\cong$ 1 mm and thickness in the range 400-450 µm) printed in this way display optical transmittance higher than 90% in the whole visible range (Figure 1c), corresponding to an attenuation loss coefficient of the order of 1 cm$^{-1}$. While a low layer thickness is beneficial for the surface quality and the optical properties of the printed structures, it might raise issues related to the photopolymerization rate. Both the presence of many interfaces between adjacent layers and the favored oxygen diffusion are expected to be detrimental for the free-radical photopolymerization process of acrylate-based pre-polymers. To investigate such aspects, we exploit the system schematized in Figure 1a for *in-situ*, real-time monitoring of photopolymerization. Figure 2a shows a typical temporal evolution of the intensity, $I_{BS}$, of the UV light backscattered by a sample of BisEMA under continuous laser irradiation. Here we focus the laser into the droplets in correspondence of their maximum height and monitor the intensity of the backscattered light. This method allows a volume nearby the droplet center, at the interface with the substrate, to be cured and probed. Due to the small size of the laser spot (1.2 μm) with respect to the droplet size (from a few to hundreds of μm), the points probed are far from the fluid-substrate contact line. $I_{BS}$ starts to increase after a few seconds following the switch-on of the UV laser (inset of Figure 2a). This time is likely needed to generate a density of free-radicals sufficient for initiating the polymerization, as observed in similar systems [49], which then proceeds for tens-hundreds of seconds until saturation of the backscattered signal is reached. Here, the variation of the refractive index of the cured pre-polymer and of the cured depth, occurring upon changing the energy per unit area delivered to the pre-polymer (the latter described by the Jacobs' cure depth relation [50]), are expected to contribute to the observed trend. To quantify the photo-polymerization time needed for curing samples with various characteristic volumes and geometries (droplets and thin films), a characteristic time, $t_{pol}$, is introduced, that is the time needed for the backscattered signal to be 1.5 time larger than the initial one: $I_{BS}(t_{pol}) = 1.5 \times I_{BS}(t = 0)$. The choice of such value for $t_{pol}$ is motivated by the observation that structures exposed for a time $\geq t_{pol}$ are structurally robust, and stable after rinsing/developing in suitable solvents. Therefore, such parameter might provide useful information for the experimental optimization of the exposure time in vat photopolymerization. The first sample studied is composed by droplets of BisEMA, with height varying in the range 1-350 μm as resulting by a substrate dewetting process (see details in Material and methods, Supplementary Figure 2, and micrographs in the insets of Figure 2b). $t_{pol}$ is about 20 s for droplets with height in the interval 20-100 μm, whereas an increase of $t_{pol}$ (by up to an order of magnitude) is measured for droplets with a height of 16 μm (Figure 2b). Moreover, for heights below 10 μm, no photopolymerization is observed, i.e. the droplets remain liquid even after long (20 minutes) UV exposures.

Similar results are obtained by *ex-situ* confocal Raman microscopy of samples made of E-shell® 600, by measuring the monomer conversion factor, $C_R$. To this aim, controlled arrays of droplets of E-shell® 600 pre-polymer are deposited by a material jetting system (Supplementary Figure 3) and exposed to UV light for various time intervals (Supplementary Figure 4). The photopolymerization of E-shell®600 occurs on timescales of the order of seconds and with much lower irradiation intensity with respect to BisEMA, because of the more effective free-radical generation of phosphine oxide photoinitiators at the 405 nm curing light [51]. Raman spectra show peaks at about 1608 cm⁻¹ and 1638 cm⁻¹, which are due to C=C aromatic ring stretching and to aliphatic C=C stretching, respectively [39] (Figure 2c and Supplementary Figure 4). Upon increasing the UV exposure time, the intensity of the peak at 1638 cm⁻¹ ($I_{1638}$) decreases while that of the peak at 1608 cm⁻¹ ($I_{1608}$) remains almost constant, as expected due to the photopolymerization [39,40].

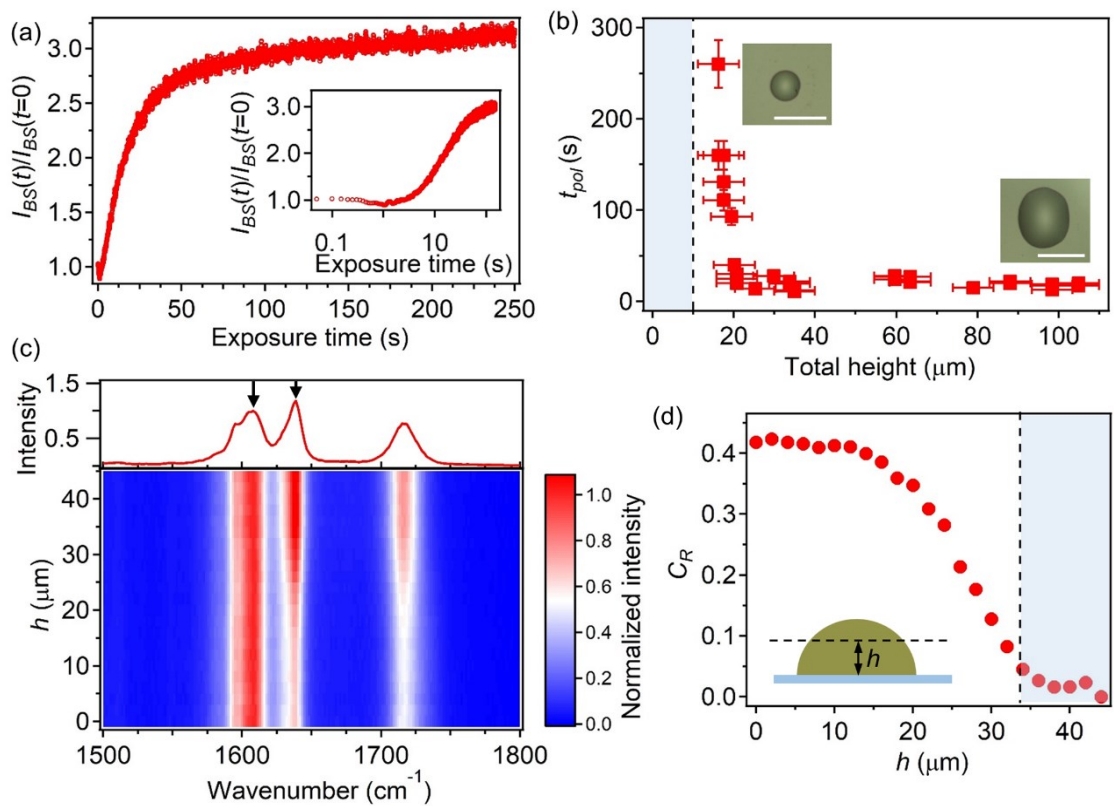

**Figure 2**. Temporal evolution of the intensity, $I_{BS}(t)$, of the UV light backscattered by the exposed region (BisEMA). Inset: $I_{BS}(t)$ at short exposure times (logarithmic temporal scale). $I_{BS}(t)$ is normalized to the signal measured at $t$=0 s. UV power density during exposure: 3×10⁷ mW cm⁻² (b) Dependence of $t_{pol}$ on the height of BisEMA droplets. The vertical dashed line highlights the droplet size below which the photopolymerization reaction is completely inhibited (shadow area). Inset: bright-field optical micrographs of BisEMA droplets. Scale bar: 100 µm. (c) Raman spectra of droplets of E-shell® resin depending on the distance, $h$, from the substrate -schematized in the inset of (d)-. An exemplary Raman spectrum is shown in the top panel. (d) Dependence of the conversion factor, $C_R$, on $h$. $C_R$ is derived from the intensities of the two modes at 1608 cm⁻¹ and 1638 cm⁻¹, highlighted by arrows in (c). As reference for calculating the $C_R$, the ratio at $h$=45 µm is considered. The dashed area shows the region close to air, showing high inhibition of free-radical polymerization.

The conversion factor, $C_R$, reaches a plateau after 7 s of UV exposure (Supplementary Figure 4). Interestingly, by measuring the spatially-resolved confocal Raman spectra at various distances, $h$, from the substrate (scheme in the inset of Figure 2d) and after 9 s of UV exposure, a monotonic dependence of $C_R$ on $h$ is found as shown in Figure 2c,d. In particular, while the droplet has an overall height of 45 µm, polymerization occurs only for $h$<35 µm ($C_R$>0.2), the 10 µm-deep layer in contact with air showing an almost null conversion factor.

Such findings can be explained by considering diffusion of oxygen, which is an efficient inhibitor of free-radical polymerization of acrylate-based pre-polymers [23]. This is in agreement with experiments that exploit such so-called *dead layers*, where polymerization is completely inhibited by oxygen, to print 3D objects in continuous runs [52]. Such effects have been also used for printing 3D micro-objects by continuous flow lithography [53]. In such works, the typical thickness of dead layers is from ∼ 1 µm to a few tens of µm, similar to the one measured here. Our methods provide *in-situ*, quantitative measurements of the layers where free-radical polymerization is inhibited, which can be highly useful for optimizing printing processes relying on a precise reaction control.

The second geometry investigated is constituted by thin films, which we made by SU-8, an epoxy resin with high mechanical stability, largely used in additive manufacturing [54-57]. In SU-8, the absorption of UV light generates reactive species, and cationic polymerization typically occurs by thermal treatment at $T>T_g$, where $T_g$ is the glass transition temperature ($T_g$∼50 °C). The intensity of the backscattered signals upon curing and the corresponding $t_{pol}$ measured for SU-8 films with various thicknesses are shown in Figure 3. A decrease of $t_{pol}$ by about a factor 2 can be appreciated for a thickness of the film above about 1.5 µm. Moreover, in Figure 3b we show the trend of the rate of the process in the regime in which the polymerization grows exponentially. We estimated $k_{pol}=1/\tau_{pol}$ (where $\tau_{pol}$ is the characteristic time derived by fitting the data to a Boltzmann function, see Supplementary Figure 5), which is almost constant for the different investigated films.

For SU-8 films, the polymerization is expected to occur mainly by a photothermal process, and the absorption of UV radiation can generate both the cationic species and the increase of temperature necessary for polymerization. By considering an incident power $P$, of the order of 1 mW (corresponding to an intensity of the order of $10^5$ W cm$^{-2}$), the expected temperature increase for thin films can be estimated as [58,59]: $\Delta T = \frac{A_{SU-8}P}{2\sqrt{\pi}w_0\Lambda_{SU-8}}$, where $A_{SU-8}$ is the fraction of light intensity absorbed by SU-8 ($A_{SU-8}$ = 2.5% at the wavelength of the incident laser, for a sample with thickness 4.5 μm, Supplementary Figure 6), $w_0$ is the spot radius of the beam used for photopolymerization (about 0.6 μm in our system) and $\Lambda_{SU-8}$ is the thermal conductivity of SU-8 ($\Lambda_{SU-8}$=0.2 W m$^{-1}$ K$^{-1}$ [60]). The dependence of the local temperature increase, $\Delta T$, on the thickness ($d_{film}$) of SU-8 comes from the absorbance of SU-8 films: $A_{SU-8} = 1 - e^{-d_{film}\alpha_{SU-8}}$, where $\alpha_{SU-8}$ is about $5\times10^3$ m$^{-1}$. A temperature increase $\Delta T$~60 °C and $\Delta T$~6 °C is calculated for $d_{film}$=4.5 μm and $d_{film}$=0.5 μm, respectively. Therefore, for thicker films the local temperature can be > 80 °C (i.e. >$T_g$), and polymerization occurs in less than a minute as observed in Figure 3a, whereas for thinner films the local temperature is lower, thus requiring the exposure times to be increased by about a factor 2 for polymerization to occur.

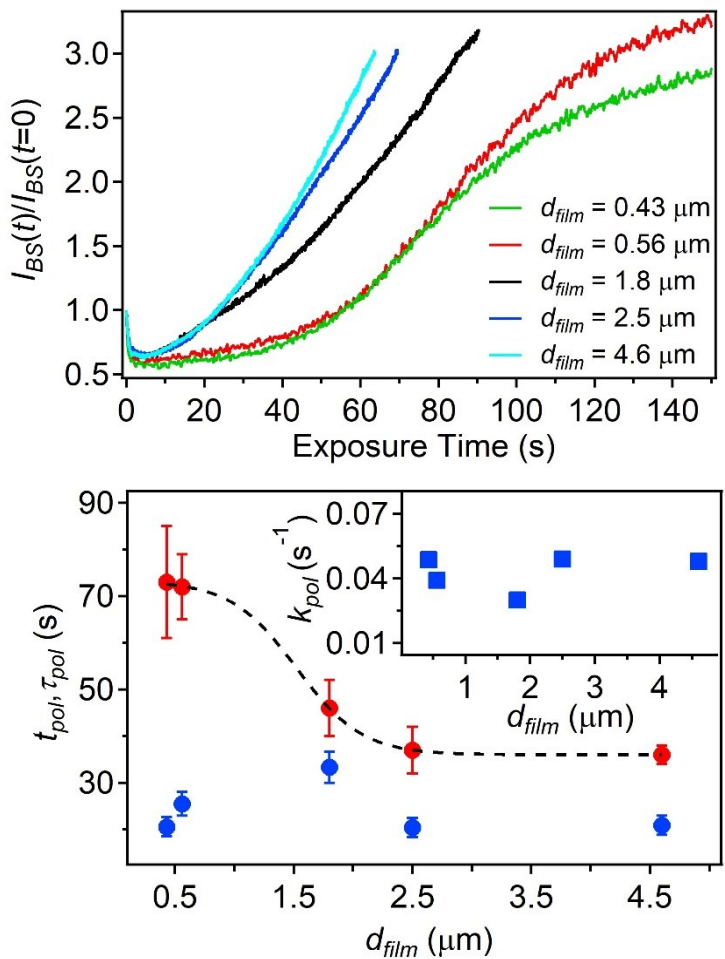

**Figure 3.** (a) Temporal evolution of the intensity, $I_{BS}(t)$, of the UV light backscattered by the exposed region (SU-8 film with different thicknesses, $d_{film}$). $I_{BS}(t)$ is normalized to the signal measured at $t = 0$. (b) Dependence of reaction times, $t_{pol}$ (red circles) and $\tau_{pol}$ (blue circles) on the film thickness. The inset shows the dependence of $k_{pol} = 1/\tau_{pol}$ on the film thickness.

## 3. Modeling of the photopolymerization process

In order to describe the size effects, we develop a phenomenological model of the photopolymerization process. First, the dependence of the probability of the activation of reactive species on the optical path length determining the overall light-polymer interaction time should be taken into account. Indeed, in thin pre-polymer layers this interaction time is short, therefore the probability of activation of reactive species is relatively low, whereas in thick films this probability is much higher. An analogous behavior has been described for polymerization activated by photothermal effects, the increase of temperature is higher for thicker samples, as discussed above.

Taking these considerations into account and following previous studies of the polymerization reactions in bulk materials similar to the one here investigated [24] the photopolymerization process is described on the basis of a simple kinetic equation:

$$\frac{d}{dt}c(t) = k(c)\left[1 - c(t)\right], \tag{1}$$

where $c$ is the polymerization degree, $t$ is time, $k$ is an effective reaction constant, for which a dependence on $c$ and on the thickness, $d_F$, of the film has been introduced phenomenologically, in order to account for the various effects here observed that might affect the photopolymerization kinetics:

$$k(c) = k_0 \left\{ 1 - \exp\left[ -\left( d_F / d(c) \right)^\alpha \right] \right\}. \tag{2}$$

In Equation (2), $k_0$ is the polymerization reaction constant in the bulk system, which is proportional to the UV beam intensity $I_0$ ($k_0 = I_0 k_{chem}$, where $k_{chem}$ is a material-dependent constant) and the normalizing thickness, $d(c)$, depends on the polymerization degree as: $d(c) = d_0 \left( c_0 / c \right)^\beta$, with $c_0 << 1$ being a polymerization degree below which no effective activation of reactive species occurs. $\alpha$ and $\beta$ are parameters which account for the specific structure of the formed polymer system and for the specific features of the polymerization reaction, respectively. The latter might include, for instance, oxygen diffusion in droplets of BisEMA and E-Shell® 600, and photothermal effects in SU-8 films. The general solution of the Equation (1) can be calculated by the following expression:

$$\int_0^{c(t)} \frac{dc}{k(c) \cdot (1 - c)} = t. \tag{3}$$

Some general features in particular regimes of polymerization can be derived. For example, if the photopolymerization occurs in thick films, or the polymerization degree is high enough ($c >> 1$), then $d_F / d(c) >> 1$, and the reaction constant, $k(c)$, neither depend on $d_F$ nor on the polymerization degree, $c$. As a result, the final stage of the photopolymerization process occurs in the classical regime, namely it shows an effective reaction constant not depending on $c$, and the process, which corresponds to the first order reaction, is described by a standard exponential kinetics: $c(t) = 1 - \exp(-k_0 t)$. On the other hand, in thin films at the initial process stage when $c << 1$, the polymerization kinetics can be dependent on the thickness of the reaction area. In fact, in this regime one has $d_F/d(c) << 1$, and the dependence of $k(c)$ on $d_F$ can be derived after an expansion in Taylor series, obtaining the following expression:

$$k(c) \approx k_0 \left[ d_F/d(c) \right]^{\alpha} = k_0 \left( d_F/d_0 \right)^{\alpha} \left( c/c_0 \right)^{\alpha\beta}. \tag{4}$$

The solution of Equation (3) gets the form:

$$c(t) = c_0 \left[ \left( k_0 t/c_0 \right) \left( d_F/d_0 \right)^{\alpha} \right]^{\frac{1}{1-\alpha\beta}}. \tag{5}$$

Overall, the above expressions are capable of accounting for a dependence of the rate of polymerization on the thickness of the reacting film at the initial stage of the process, while such dependence is smoothed as the thickness increases. With the polymerization reaction flowing, the thickness dependence of the rate of polymerization becomes weaker, and at the final stage of the process it will be almost negligible. The cross-over from the initial (size dependent) regime to the final one occurs at the time: $t_{Po} \approx \left( c_0/k_0 \right) \left( d_0/d_F \right)^{1/\beta}$, after which the polymerization reaction will occur according to standard kinetics described by an exponential decay.

## 4. Discussion

The phenomenological model introduced above, allows the overall kinetics of photopolymerization to be calculated by using Equation (3), varying the characteristic size parameter of the cured material. The results for different film thicknesses are shown in the Figure 4a.

The comparison of the curves of Figure 4a with the experimental data (Figure 3a), evidence some differences at the initial stage of the process. More specifically, while the curves of Figure 4a increase monotonically with time, the experimental curves show a non-monotonical behaviour.

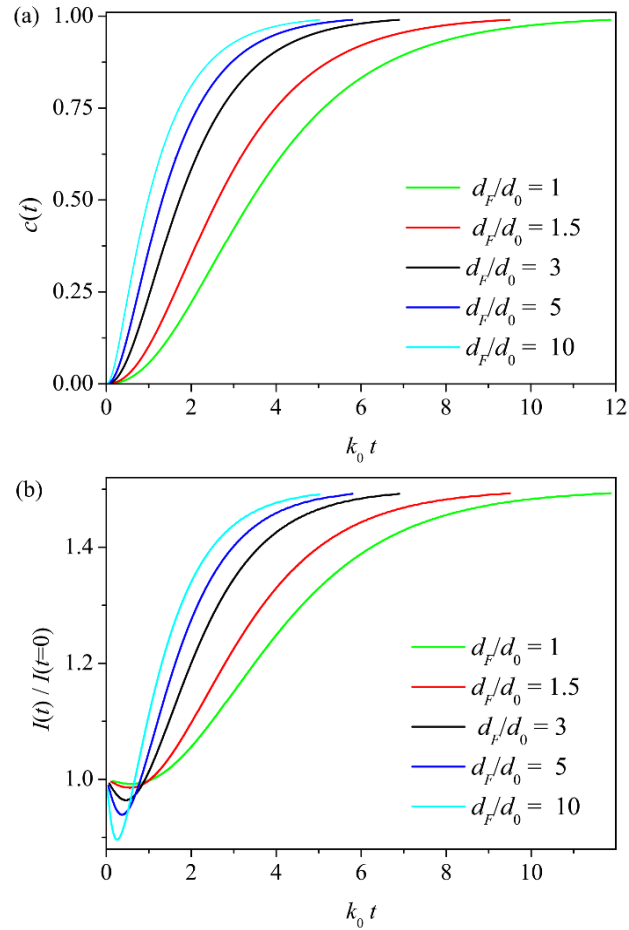

**Figure 4**. (a) Photopolymerization kinetics calculated for different film thicknesses ($d_F$). (b) Calculated temporal evolution of intensity of the backscattered light, $I(t)$, for different thicknesses ($d_F$). Other parameters: $\alpha = 0.75$, $\beta = 0.75$, $c_0 = 2$, and $\kappa = 0.5$ and $\gamma = 0.25$.

This feature in experimental data can be attributed to the used procedure, which relies on measurements of the intensity of the UV-beam activating the photopolymerization and backscattered by the polymerized region. In fact, such measured intensity of the backscattered light is determined by two factors, namely (i) its increase for increasing conversion degrees, which can be summarized by the following expression: $I_1(t)=1+\kappa c(t)$; and (ii) a contribution describing the beam intensity attenuation due to the light absorption of species in the reacting film: $I_2(t)=1-\gamma\, dc(t)/dt$. Overall, the calculated backscattered intensity can be written as:

$$I(t) = I_1(t)I_2(t) = [1 + \kappa c(t)][1 - \gamma dc(t)/dt] \qquad (6)$$

The temporal evolution of $I(t)$ for various thickness values is shown in Figure 4b. Notably, the calculated curves display a trend of non-monotonic kinetics that well reproduces the measured one ($I_{BS}$ in Figure 2a and 3a), evidencing the effectiveness of the developed model for describing photopolymerization processes taking into account potential size effects, independently on the specific nature of the underlying curing mechanism.

## 5. Conclusions

The kinetics of photopolymerization reactions are affected by the restricted volume of pre-polymer layers. An *in-situ* method for process monitoring has been developed based on real-time measurements of the light backscattered by a photo-exposed region. An increase of the time needed for the polymerization by UV light is found as the thickness of the film is decreased below about 1.5 or 10 μm, for materials relying on cationic/photothermal and on free-radical polymerization, respectively. A phenomenological model for photopolymerization has been developed, which is capable of reproducing the trend measured by *in-situ* monitoring. These results show that while a decrease of the layer thickness is effective for printing 3D objects with improved surface quality, the associated increase of the curing time demands a proper choice of compensating polymerization parameters. Strategies might include the increase of photoinitiator concentration in used materials as well as the increase of light intensity during printing [23,52,53]. In this framework, our study provides an effective experimental tool and theoretical framework for engineering 3D printing based on UV photopolymerization, for improving the achievable spatial resolution, and for enhancing additive manufacturing with composite and heterogeneous materials.

**Acknowledgments**

The research leading to these results has received funding from the European Research Council (ERC) under the European Union's Horizon 2020 research and innovation programme (grant agreement No. 682157, "*x*PRINT") and from the Italian Minister of University and Research PRIN 2017PHRM8X project ("3D-Phys"). S. Uttiya is acknowledged for preparation of BisEMA samples, F. Cardarelli for support with confocal microcopy and L. Sercia for SEM imaging.

**Appendix A. Supplementary data**

Supplementary material related to this article can be found in the online version.

**References**

[1] J. Deng, L. Wang, L. Liu, W. Yang, Developments and new applications of UV-induced surface graft polymerizations, Prog. Polym. Scie. 34 (2009) 156-193, https://doi.org/10.1016/j.progpolymsci.2008.06.002.

[2] J. G. Leprince, W. M. Palin, M. A. Hadis, J. Devaux, G. Leloup, Progress in dimethacrylate-based dental composite technology and curing efficiency, Dent. Mater. 29 (2013) 139-156, https://doi.org/10.1016/j.dental.2012.11.005.

[3] S. Waheed, J. M. Cabot, N. P. Macdonald, T. Lewis, R. M. Guijt, B. Paull, M. C. Breadmore, 3D printed microfluidic devices: enablers and barriers, Lab Chip 16 (2016) 1993-2013, https://doi.org/10.1039/C6LC00284F.

[4] K. Nakamura, Photopolymers: Photoresist Materials, Processes, and Applications, CRC Press, Boca Renton, 2015.

[5] R. Gurarslan, A. E. Tonelli, Do we need to know and can we determine the complete macrostructures of synthetic polymers?, Prog. Polym. Scie. 65 (2017) 42-52, https://doi.org/10.1016/j.progpolymsci.2016.09.001.

[6] C. Mack, Fundamental Principles of Optical Lithography: The Science of Microfabrication, John Wiley and Sons, Chichester, West Sussex, 2007.

[7] C. Sun, N. Fang, D. M. Wu, X. Zhang, Projection micro-stereolithography using digital micro-mirror dynamic mask, Sens. Actuators A *121* (2005) 113-120, https://doi.org/10.1016/j.sna.2004.12.011.

[8] X. Zheng, J. Deotte, M. P. Alonso, G. R. Farquar, T. H. Weisgraber, S. Gemberling, H. Lee, N. Fang, C. M. Spadaccini, Design and optimization of a light-emitting diode projection micro-

stereolithography three-dimensional manufacturing system, Rev. Sci. Instrum. 83 (2012) 125001, https://doi.org/10.1063/1.4769050.

[9] J. Chen, Y. Wu, X. Li, F. Cao, Y. Gu, K. Liu, X. Liu, Y. Dong, J. Ji, H. Zeng, Simple and fast patterning process by laser direct writing for perovskite quantum dots, Adv. Mater. Technol. 2 (2017) 1700132, https://doi.org/10.1002/admt.201700132.

[10] R. W. Boyd, Nonlinear Optics, 3rd Ed., Academic Press, New York, 2008.

[11] S. Kawata, H.-B. Sun, T. Tanaka, K. Takada, Finer features for functional microdevices, Nature 412 (2001) 697-698, https://doi.org/10.1038/35089130.

[12] L. Li, R. R. Gattass, E. Gershgoren, H. Hwang, J. T. Fourkas, Achieving $\lambda/20$ resolution by one-color initiation and deactivation of polymerization, Science 324 (2009) 910-913, https://doi.org/10.1126/science.1168996.

[13] T. F. Scott, B. A. Kowalski, A. C. Sullivan, C. N. Bowman, R. R. McLeod, Two-color single-photon photoinitiation and photoinhibition for subdiffraction photolithography, Science 324 (2009) 913-917, https://doi.org/10.1126/science.1167610.

[14] T. L. Andrew, H.-Y. Tsai, R. Menon, Confining light to deep subwavelength dimensions to enable optical nanopatterning, Science 324 (2009) 917-921, https://doi.org/10.1126/science.1167704.

[15] F. Zhang, L. Zhu, Z. Li, S. Wang, J. Shi, W. Tang, N. Li, J. Yang, The recent development of vat photopolymerization: A review, Addit. Manuf. 48 (2021) 102423, https://doi.org/10.1016/j.addma.2021.102423.

[16] A. Camposeo, L. Persano, M. Farsari, D. Pisignano, Additive manufacturing: applications and directions in photonics and optoelectronics, Adv. Optical Mater. 7 (2019) 1800419, https://doi.org/10.1002/adom.201800419.

[17] J. Fischer, M. Wegener, Three-dimensional optical laser lithography beyond the diffraction limit, Laser Photonics Rev. 7 (2013) 22, https://doi.org/10.1002/lpor.201100046.

[18] M. Regehly, Y. Garmshausen, M. Reuter, N. F. König, E. Israel, D. P. Kelly, C.-Y. Chou, K. Koch, B. Asfari, S. Hecht, Xolography for linear volumetric 3D printing, Nature 588 (2020) 620-624, https://doi.org/10.1038/s41586-020-3029-7.

[19] D. M. Hartmann, O. Kibar, S. C. Esener, Characterization of a polymer microlens fabricated by use of the hydrophobic effect, Opt. Lett. 25 (2000) 975-977, https://doi.org/10.1364/OL.25.000975.

[20] Y. Peng, X. Guo, R. Liang, Y. Mou, H. Cheng, M. Chen, S. Liu, Fabrication of microlens arrays with controlled curvature by micromolding water condensing based porous films for deep ultraviolet LEDs, ACS Photonics 4 (2017) 2479-2485, https://doi.org/10.1021/acsphotonics.7b00692.

[21] R. Magazine, B. van Bockove, S. Borandeh, J. Seppälä, 3D inkjet-printing of photo-crosslinkable resins for microlens fabrication, Addit. Manuf. (2021), https://doi.org/10.1016/j.addma.2021.102534.

[22] V. V. Krongauz, E. R. Schmelzer, R. M. Yohannan, Kinetics of anisotropic photopolymerization in polymer matrix, Polymer 32 (1991) 1654-1662, https://doi.org/10.1016/0032-3861(91)90402-5.

[23] S. Clark Ligon, B. Husár, H. Wutzel, R. Holman, R. Liska, Strategies to reduce oxygen inhibition in photoinduced polymerization, Chem. Rev. 114 (2014) 557, https://doi.org/10.1021/cr3005197.

[24] C. S. Reddy, A. Arinstein, E. Zussman, Polymerization kinetics under confinement, Polym. Chem. 2 (2011) 835-839, https://doi.org/10.1039/C0PY00285B.

[25] R. D. Farahani, M. Dubé, D. Therriault, Three-dimensional printing of multifunctional nanocomposites: manufacturing techniques and applications, Adv. Mater. 28 (2016) 5794-5821, https://doi.org/10.1002/adma.201506215.

[26] S. Biria, I. D. Hosein, Control of morphology in polymer blends through light self-trapping: an in situ study of structure evolution, reaction kinetics, and phase separation, Macromolecules 50 (2017) 3617-3626, https://doi.org/10.1021/acs.macromol.7b00484.

[27] X. Zhao, D. W. Rosen, An implementation of real-time feedback control of cured part height in Exposure Controlled Projection Lithography with in-situ interferometric measurement feedback, Addit. Manuf. 23 (2018) 253-263, https://doi.org/10.1016/j.addma.2018.07.016.

[28] T. Hafkamp, G. van Baars, B. de Jager, P. Etman, Real-time feedback controlled conversion in vat photopolymerization of ceramics: A proof of principle, Addit. Manuf. 30 (2019) 100775, https://doi.org/10.1016/j.addma.2019.06.026.

[29] J. B. Mueller, J. Fischer, F. Mayer, M. Kadic, M. Wegener, Polymerization kinetics in three-dimensional direct laser writing, Adv. Mater. 26 (2014) 6566-6571, https://doi.org/10.1002/adma.201402366.

[30] C. A. Bonino, J. E. Samorezov, O. Jeon, E. Alsberg, S. A. Khan, Real-time in situ rheology of alginate hydrogel photocrosslinking, Soft Matter 7 (2011) 11510-11517, https://doi.org/10.1039/C1SM06109G.

[31] A. Cusano, G. Breglio, M. Giordano, A. Calabrò, A. Cutolo, L. Nicolais, An optoelectronic sensor for cure monitoring in thermoset-based composites, Sens. Actuators 84 (2000) 270-275, https://doi.org/10.1016/S0924-4247(00)00361-7.

[32] K. Classens, T. Hafkamp, S. Westbeek, J. J. C. Remmers, S. Weiland, Multiphysical modeling and optimal control of material properties for photopolymerization processes, Addit. Manuf. 38 (2021) 101520, https://doi.org/10.1016/j.addma.2020.101520.

[33] L. Wommer, P. Meiers, I. Kockler, R. Ulber, P. Kampeis, Development of a 3D-printed single-use separation chamber for use in mRNA-based vaccine production with magnetic microparticles, Eng. Life Sci. 21 (2021) 573-588, https://doi.org/10.1002/elsc.202000120.

[34] S. K. Rath, F. Y. C. Boey, M. J. M. Abadie, Cationic electron-beam curing of a high-functionality epoxy: effect of post-curing on glass transition and conversion. Polym. Int. 53 (2004) 857–862, https://doi.org/10.1002/pi.1383.

[35] T. L. Tan, D. Wong, P. Lee, R. S. Rawat, A. Patran, Study of a chemically amplified resist for X-ray lithography by Fourier transform infrared spectroscopy, Appl. Spectrosc. 58 (2004) 1288-1294, https://doi.org/10.1366/0003702042475402.

[36] M. T. Do, T. T. N. Nguyen, Q. Li, H. Benisty, I. Ledoux-Rak, N. D. Lai, Submicrometer 3D structures fabrication enabled by one-photon absorption direct laser writing, Opt. Express 21 (2013) 20964-20973, https://doi.org/10.1364/OE.21.020964.

[37] M. Nordström, D. A. Zauner, A. Boisen, J. Hübner, Single-mode waveguides with SU-8 polymer core and cladding for MOEMS applications, J. Light. Technol. 25 (2007) 1284-1289, https://doi.org/10.1109/JLT.2007.893902.

[38] F. Aloui, L. Lecamp, P. Lebaudy, F, Burel, Relationships between refractive index change and light scattering during photopolymerization of acrylic composite formulations, J. Eur. Ceram. Soc. 36 (2016) 1805-1809, https://doi.org/10.1016/j.jeurceramsoc.2016.01.033.

[39] M. Par, N. Spanovic, T. T. Tauböck, T. Attin, Z. Tarle, Degree of conversion of experimental resin composites containing bioactive glass 45S5: the effect of post-cure heating, Sci. Rep. 9 (2019) 17245, https://doi.org/10.1038/s41598-019-54035-y.

[40] Y. Cai, J. L. P. Jessop, Decreased oxygen inhibition in photopolymerized acrylate/epoxide hybrid polymer coatings as demonstrated by Raman spectroscopy, Polymer 47 (2006) 6560-6566, https://doi.org/10.1016/j.polymer.2006.07.031.

[41] M. Born, E. Wolf, Principles of Optics, Sixth Ed., Pergamon Press, Oxford, 1980.

[42] Refractive index of BisEMA is provided by the supplier:

https://www.sigmaaldrich.com/catalog/product/aldrich/455059?lang=it®ion=IT. Accessed May 2022.

[43] Y. Pan, X. Zhao, C. Zhou, Y. Chen, Smooth surface fabrication in mask projection based stereolithography, J. Manuf. Processes 14 (2012) 460-470, https://doi.org/10.1016/j.jmapro.2012.09.003.

[44] A. Heinrich, M. Rank, P. Maillard, A. Suckow, Y. Bauckhage, P. Rößler, J. Lang, F. Shariff, S. Pekrul, Additive manufacturing of optical components, Adv. Opt. Technol. 5 (2016) 293-301, https://doi.org/10.1515/aot-2016-0021.

[45] X. Chen, W. Liu, B. Dong, J. Lee, H. O. T. Ware, H. F. Zhang, C. Sun, High-speed 3D printing of millimeter-size customized aspheric imaging lenses with sub 7 nm surface roughness, Adv. Mater. 30 (2018) 1705683, https://doi.org/10.1002/adma.201705683.

[46] N. Vaidya, O. Solgaard, 3D printed optics with nanometer scale surface roughness, Microsyst. Nanoeng. 4 (2018) 18, https://doi.org/10.1038/s41378-018-0015-4.

[47] G. Shao, R. Hai, C. Sun, 3D printing customized optical lens in minutes, Adv. Optical Mater. 8 (2020) 1901646, https://doi.org/10.1002/adom.201901646.

[48] K. Kowsari, B. Zhang, S. Panjwani, Z. Chen, H. Hingorani, S. Akbari, N. X. Fang, Qi Ge, Photopolymer formulation to minimize feature size, surface roughness, and stair-stepping in digital light processing-based three-dimensional printing, Addit. Manuf. 24 (2018) 627-638, https://doi.org/10.1016/j.addma.2018.10.037.

[49] S. Wu, M. Straub, M. Gu, Single-monomer acrylate-based resin for three-dimensional photonic crystal fabrication, Polymer 46 (2005) 10246-10255, https://doi.org/10.1016/j.polymer.2005.08.030.

[50] P. F. Jacobs, Rapid prototyping and manufacturing: Fundamentals of stereolithography. Society of manufacturing engineers, Dearborn, 1992.

[51] S. C. Ligon, R. Liska, J. Stampfl, M. Gurr, R. Mülhaupt, Polymers for 3D printing and customized additive manufacturing, Chem. Rev. 117 (2017) 10212–10290, https://doi.org/10.1021/acs.chemrev.7b00074.

[52] J. R. Tumbleston, D. Shirvanyants, N. Ermoshkin, R. Janusziewicz, A. R. Johnson, D. Kelly, K. Chen, R. Pinschmidt, J. P. Rolland, A. Ermoshkin, E. T. Samulski, J. M. DeSimone, Continuous liquid interface production of 3D objects, Science 347 (2015) 1349-1352, https://doi.org/10.1126/science.aaa2397.

[53] D. Dendukuri, P. Panda, R. Haghgooie, J. M. Kim, T. A. Hatton, P. S. Doyle, Modeling of oxygen-inhibited free radical photopolymerization in a PDMS microfluidic device, Macromolecules 41 (2008) 8547-8556, https://doi.org/10.1021/ma801219w.

[54] W. H. Teh, U. Dürig, G. Salis, R. Harbers, U. Drechsler, R. F. Mahrt, C. G. Smith, H.-J. Güntherodt, SU-8 for real three-dimensional subdiffraction-limit two-photon microfabrication, Appl. Phys. Lett. 84 (2004) 4095-4097, https://doi.org/10.1063/1.1753059.

[55] J.-W. Su, X. Tao, H. Deng, C. Zhang, S. Jiang, Y. Lina, J. Lin, 4D printing of a self-morphing polymer driven by a swellable guest medium, Soft Matter 14 (2018) 765-772, https://doi.org/10.1039/C7SM01796K.

[56] W. Chu, Y. Tan, P. Wang, J. Xu, W. Li, J. Qi, Y. Cheng, Centimeter-height 3D printing with femtosecond laser two-photon polymerization, Adv. Mater. Technol. 3 (2018) 1700396, https://doi.org/10.1002/admt.201700396.

[57] H. Yang, Y.-Y. Zhao, M.-L. Zheng, F. Jin, X.-Z. Dong, X.-M. Duan, Z.-S. Zhao, Stepwise optimized 3D printing of arbitrary 3D structures at millimeter scale with high precision surface, Macromol. Mater. Eng. 304 (2019) 1900400, https://doi.org/10.1002/mame.201900400.

[58] D. C. Cahill, Analysis of heat flow in layered structures for time-domain thermoreflectance, Rev. Sci. Instrum. 75 (2004) 5119-5122, https://doi.org/10.1063/1.1819431.

[59] J. L. Braun, C. J. Szwejkowski, A. Giri, P. E. Hopkins, On the steady-state temperature rise during laser heating of multilayer thin films in optical pump–probe techniques, J. Heat Transfer 140 (2018) 052801, https://doi.org/10.1115/1.4038713.

[60] S. H. Oh, K.-C. Lee, J. Chun, M. Kim, S. S. Lee, Micro heat flux sensor using copper electroplating in SU-8 microstructures, J. Micromech. Microeng. 11 (2001) 221-225, https://doi.org/10.1088/0960-1317/11/3/310.

## Supplementary Material

## Impact of size effects on photopolymerization and its optical monitoring *in-situ*

Andrea Camposeo,[a,*] Aristein Arkadii,[b] Luigi Romano,[a] Francesca D'Elia,[c] Filippo Fabbri,[a]

Eyal Zussman,[b] Dario Pisignano[a,d]

*[a]NEST, Istituto Nanoscienze-CNR and Scuola Normale Superiore, Piazza S. Silvestro 12, I-56127*

*Pisa, Italy*

*[b]Department of Mechanical Engineering, Technion-Israel Institute of Technology, Haifa, 32000,*

*Israel*

*[c]NEST, Scuola Normale Superiore, Piazza San Silvestro 12, I-56127 Pisa, Italy*

*[d]Dipartimento di Fisica, Università di Pisa, Largo B. Pontecorvo 3, I-56127 Pisa, Italy*

*\*Corresponding author: e-mail: andrea.camposeo@nano.cnr.it*

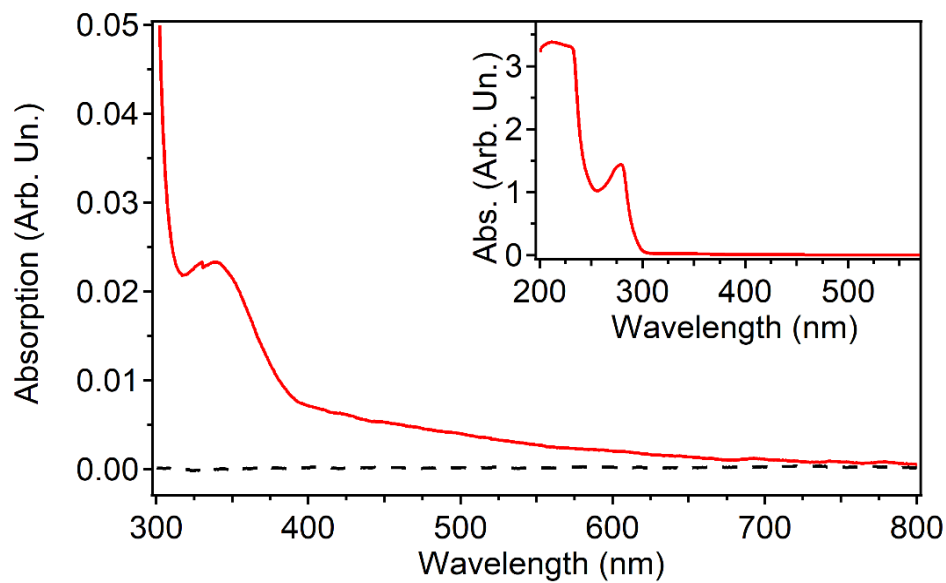

**Supplementary Figure 1**. Absorption spectra of the bisphenol-A-ethoxylate dimethacrylate/2,2-dimethoxy-2-phenylacetophenone mixture (2% weigth:weight, red continuous line) and of a reference quartz substrate (black dashed line). The inset shows the spectra of the mixture in the UV-visible wavelength interval.

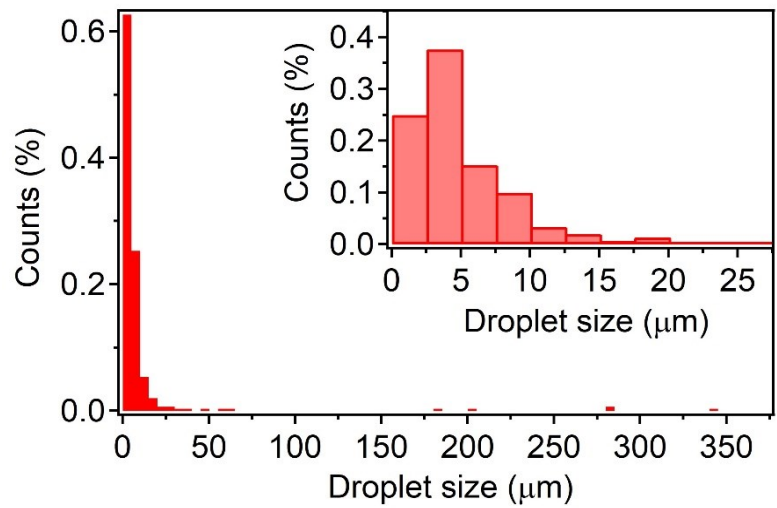

**Supplementary Figure 2**. Distribution of the size (diameter) of the droplets of BisEMA. The inset shows a magnification of the size distribution in the range 0-30 μm.

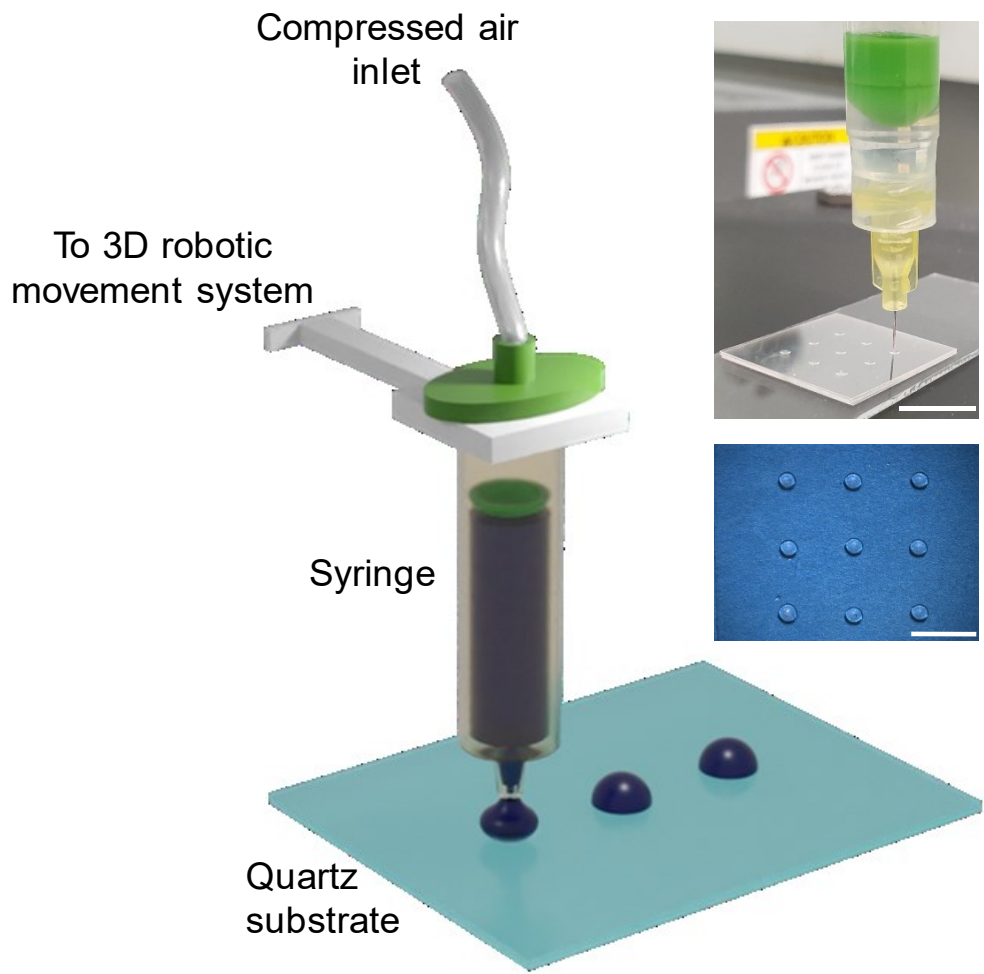

**Supplementary Figure 3**. Schematic illustration of the material jetting system used for printing the droplets of E-Shell® 600. The top inset shows a photographs of the needle used for the extrusion. Scale bar: 1 cm. The bottom inset shows a 3×3 array of printed droplets. Scale bar: 2 mm.

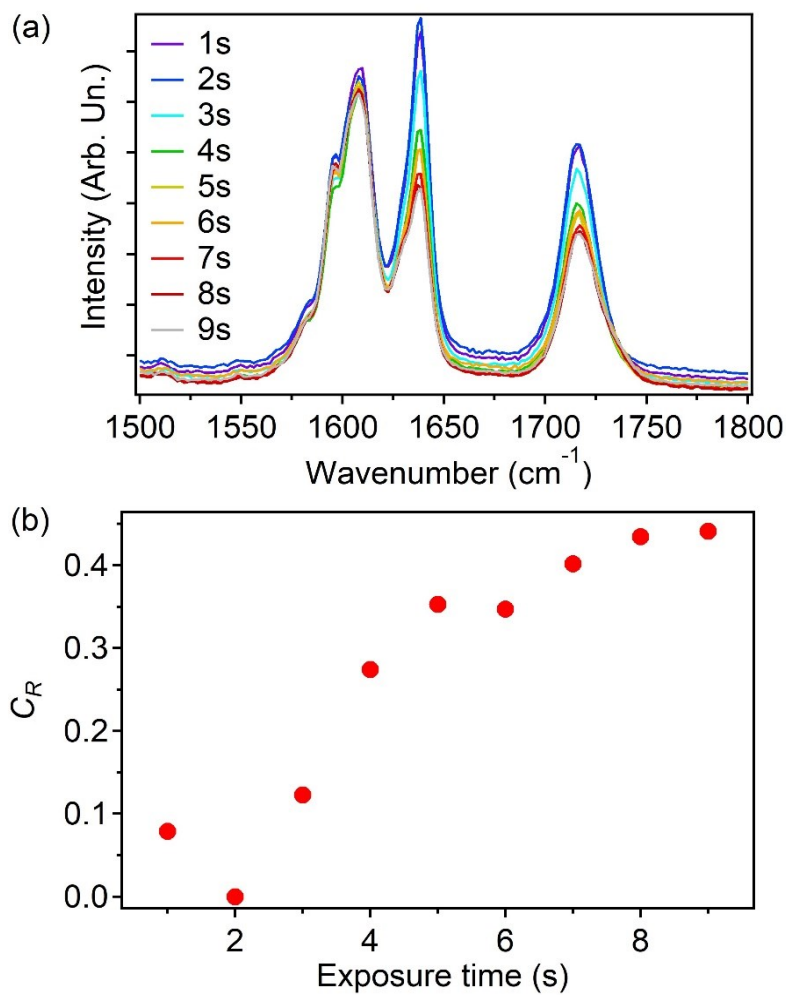

**Supplementary Figure 4**. Analysis of the photopolymerization of droplets of E-Shell® 600, printed by material jetting. (a) Raman spectra for various exposure times to the UV light. (b) Dependence of the conversion factor, $C_R$, on the UV exposure time. UV power density during exposure: 3 mW cm$^{-2}$.

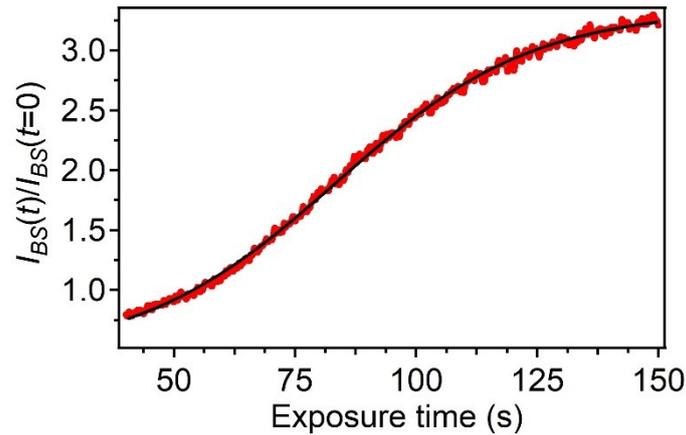

**Supplementary Figure 5**. Temporal evolution of the intensity, $I_{BS}(t)$, of the UV light backscattered by the exposed region for a SU-8 film with thickness, $d_{film}$=0.56 μm. The black line is a fit to the data by a Boltzmann-like expression that can account for both the exponential temporal evolution of the photopolymerization at the early stage and the saturation regime occurring at exposure times >100 s: $\frac{I_{BS}(t)}{I_{BS}(t=0)} = A_2 + \frac{A_1 - A_2}{1 + \exp[(t - t_{p0})/\tau_{pol}]}$. Here, $A_1$ and $A_2$ are constants and $\tau_{pol}$ is the characteristic time of the regime in which the photopolymerization grows exponentially.

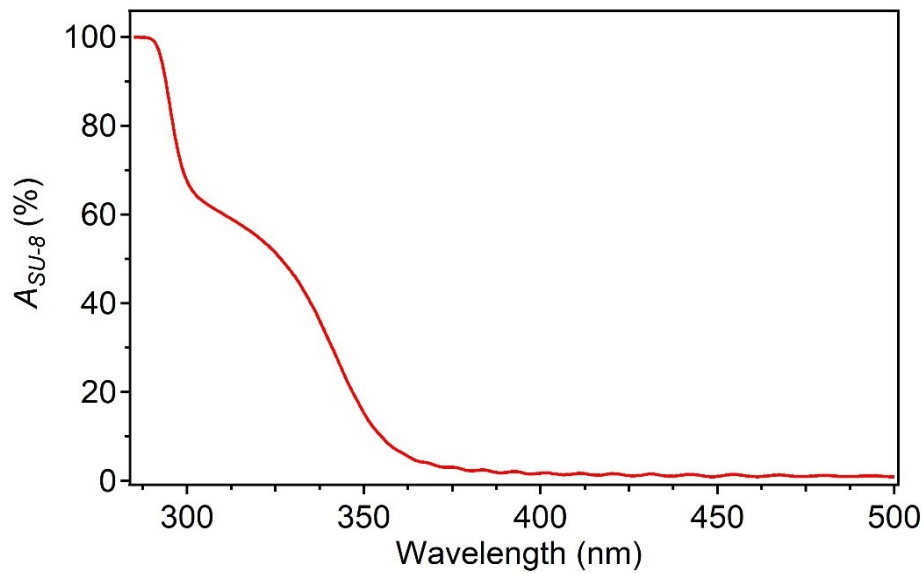

**Supplementary Figure 6**. Spectrum of the fraction of light intensity absorbed of a film of SU-8, $A_{SU\text{-}8}$ (thickness = 4.5 μm).